\documentclass{LT23auth}
\include{psfig}
\usepackage{epsfig}

\begin{document}

\begin{frontmatter}

\title{Entangled Hanbury Brown Twiss effects with edge states}

\author{M. B\"uttiker$^1$, P. Samuelsson,E. V. Sukhorukov},

\address{D\'epartement de Physique Th\'eorique, 
Universit\'e de Gen\`eve, CH-1211 Gen\`eve 4, Switzerland.}

\thanks{Corresponding author. \\{\it E-mail adress}:buttiker@serifos.unige.ch}

\begin{abstract}
Electronic Hanbury Brown Twiss correlations are discussed for
geometries in which transport is along adiabatically guided edge
channels. We briefly discuss partition noise experiments and discuss
the effect of inelastic scattering and dephasing on current
correlations. We then consider a two-source Hanbury Brown Twiss
experiment which demonstrates strikingly that even in geometries
without an Aharonov-Bohm effect in the conductance matrix
(second-order interference), correlation functions can (due to
fourth-order interference) be sensitive to a flux.  Interestingly we
find that this fourth-order interference effect is closely related to
orbital entanglement. The entanglement can be detected via violation
of a Bell Inequality in this geometry even so particles emanate from
uncorrelated sources.
\end{abstract}

\end{frontmatter}

\section{Introduction}
In this article we are concerned with dynamical current fluctuations
(noise) in the quantized Hall regime. In particular we want to discuss
a series of experiments for electrons which are close electronic
analogs of experiments in quantum optics. Two aspects make the
quantized Hall effect \cite{vK} (QHE) particularly suitable for such a
development: First, the chiral nature of edge states permits transport
of electrons over (electronically) large distances. Not only is
backscattering suppressed but a lateral dilution of the "electron
beam" is also prevented.

The second element needed to mimick optical geometries, the
half-silvered mirror, is similarly available in the form of quantum
point contacts \cite{qpc1,qpc2} (QPC's) or gates \cite{gate}.  Indeed
in high magnetic fields a QPC permits the separate measurement of
transmitted and reflected carriers \cite{mb90}.

The quantities of interest are noise correlations between the current
fluctuations measured at two contacts of a mesoscopic conductor.  The
optical analog of this quantity is an intensity-intensity correlation
measured with two detectors. Intensity correlations of photons became
of interest with the invention by Hanbury Brown and Twiss (HBT) of an
interferometer which permitted to determine the angular diameter of
visual stars \cite{hbt}.  For photons emitted by a thermal source a
classical wave field explanation is possible. A quantum theory was put
forth by Purcell \cite{pruc}. The HBT effect contains two important
distinct but
\begin{figure}[h]
\centerline{\psfig{figure=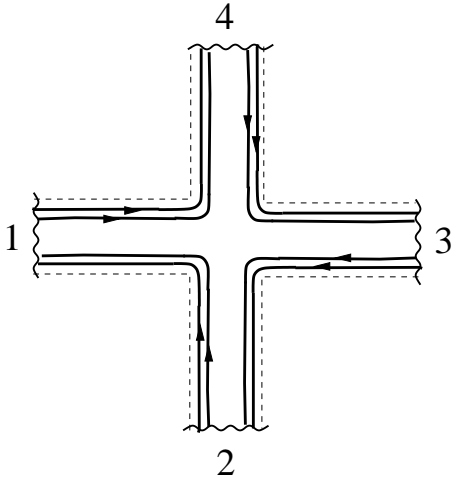,width=6.0cm}}
\caption{Cross-geometry for quantized Hall effect measurements.}
\label{fig1}
\end{figure}
fundamentally interrelated effects \cite{sam2}: First, light from
different completely uncorrelated portions of the star gives rise to
an interference effect which is visible in intensity correlations but
not in the intensities themselves.  This is a property of two particle
exchange amplitudes.  Second there is a direct statistical effect
since photons bunch whereas Fermions anti-bunch.  It was long a dream
to realize the electronic equivalent of the optical HBT
experiment. This is difficult to achieve with field emission of
electrons into vacuum because the effect is quadratic in the
occupation numbers.  This difficulty is absent in electrical
conductors where at low temperatures a Fermi gas is completely
degenerate. Initial experiments demonstrating fermionic anti-bunching
were reported by Henny et al. \cite{henny} and Oberholzer et
al. \cite{ober} using edge states and at zero-magnetic field by Oliver
et al. \cite{oliv}. Only very recently was a first experiment with a
field emission source successful \cite{vacuum}. In contrast, to date,
there is no experimental demonstration of a two particle interference
effect with electrons.

We first briefly review some basic aspects of edge state transport. We
discuss the first experiments on current-current correlations and
examine the effect of inelastic scattering and potential fluctuations
on these correlations.  We extend a Mach-Zehnder geometry to
investigate a two-source HBT set-up in which the conductance is
phase-insensitive but the current correlations are sensitive to phases
accumulated along edge states \cite{sam2}. We analyze entanglement in
this geometry with a Bell inequality \cite{sam2}.
\section{Edge States and the quantized Hall effect}
Edge states are quantized skipping orbit states. Although edge states
were discussed very soon after the discovery of the QHE by Halperin
\cite{halp}, for a considerable time, they were seemingly irrelevant
for the description of experiments. After all, in a macroscopic Hall
bar, the contribution of edge states to the total density of states is
of measure zero. It was only the increasing concern with mesoscopic
physics that eventually brought about a new look at the QHE and lifted
the notion of edge states from a theoretical concept to one that could
be experimentally tested. For this advance it was necessary to
understand the role of contacts as current injectors and absorbers of
carriers with the possibility to generate and measure
non-equilibrium (or selective) populations of edge states
\cite{mb88,noneq,nerev}.

The simplest (ideal) geometry is shown in Fig. \ref{fig1}. For a Fermi
energy between the N-th and N+1-th bulk Landau level N edge states
follow the boundary of the sample. Their significance is in the fact
that they are the only states at the Fermi energy which connect
contacts \cite{mb88}.  Each edge state provides a quantum channel
which permits transmission of carriers with unit probability from the
metallic contact from which the edge state emerges to the metallic
contact which follows it clockwise on the perimeter of the sample.
Since the conductance $G_{\alpha\beta} = I_{\alpha}/V_{\beta}$ is
equal to (minus) the sum of all transmission probabilities we have
$G_{41} = G_{34} = G_{23} = G_{12} = - (e^{2}/h) N$.  All other
elements of the conductance matrix vanish.  Taking into account that
the measured resistance is $R_{\alpha\beta,\gamma\delta} = (V_{\gamma}
- V_{\delta})/I(\alpha \Rightarrow \beta )$ where the first pair of
indices indicates the carrier source and sink contact and the second
pair denotes the voltage probes, one easily finds that for the
conductor of Fig. \ref{fig1} Hall resistances of the type $R_{13,42}$
are quantized and given by $R_{13,42} = (h/e^{2})(1/N)$ whereas
longitudinal resistances of the type $R_{14,23}$ vanish \cite{mb88}.

This discussion treats current contacts and voltage contacts on equal
footing.  All conductances are evaluated at the Fermi energy. On the
other hand the above discussion can not be used to find the current
densities inside the sample: like true charge densities are found only
with help of Poisson's equations the true current distribution must be
found from a self-consistent analysis \cite{tcmb} (which determines
the Hall potential).
\section{Fundamentals of noise}
Fundamentally there are only two sources of noise \cite{mb92,ybmb}:
First, thermal agitation of carriers in the contacts leads to
fluctuations in the occupation number of incident states and gives
rise to Nyquist-Johnson noise. Second, a quantum state which has more
than one final state generates partition noise. We will briefly
discuss these two sources of noise.

The average occupation number of a state in contact $\alpha$ is given
by the Fermi distribution function $f_{\alpha} =
(exp((E-\mu_{\alpha})/kT)+1)^{-1}$ where $\mu_{\alpha}$ is the
(electro) chemical potential of the contact.  The average occupation
number is $\langle n_{\alpha} \rangle = f_{\alpha}$. Fluctuations
$\Delta n_{\alpha} = n_{\alpha} - \langle n_{\alpha} \rangle $ away
from the average are characterized by mean square fluctuations $
\langle (\Delta n_{\alpha})^{2} \rangle = f_{\alpha}(1- f_{\alpha})$
determined by the derivative of the Fermi function $f_{\alpha}(1-
f_{\alpha}) = -kT (df_{\alpha}/dE)$.

If all contacts are at the same potential this determines the
Johnson-Nyquist noise.  In particular for the zero-frequency noise
power spectrum $S_{\alpha\beta}$ of the current fluctuations at
contact $\alpha$ and $\beta$ defined through
\begin{equation}
S_{\alpha\beta} = 2\int dt \langle \Delta I_{\alpha}(t)\Delta I_{\beta}(0) \rangle,
\end{equation}
where $\Delta I_{\alpha}(t) = I_{\alpha}(t)-  \langle {I_{\alpha}} \rangle $, 
we find \cite{mb90,mb92} 
\begin{equation}
S_{\alpha\beta} = 2 kT (G_{\alpha\beta} + G_{\beta\alpha})
\end{equation}
The fluctuation dissipation theorem relates the mean squared current
fluctuations $\alpha = \beta$ to the diagonal elements of the
conductance matrix and relates the current correlations $\alpha \ne
\beta$ to the symmetrized off- diagonal elements of the conductance
matrix. An experimental test of this relation is reported in
Ref. \cite{henny}.  Similarly, voltage fluctuations are connected to
the four-terminal resistances introduced above \cite{mb90,mb92}.  The
interest in equilibrium noise is, however, limited since we obtain the
same information as from conductance or resistance measurements.

Quantum partition noise is a second fundamental source of noise. This
noise arises whenever a quantum state has more than one possible final
outcome.  In contrast to the equilibrium noise, quantum partition
noise is non-vanishing even in the zero-temperature limit.  To explain
this source of noise consider for a moment a Gedanken experiment: in
each trial a particle approaches a tunnel barrier characterized by a
transmission probability $T$ and a reflection probability $R$. We
assume that we can detect whether the particle has been reflected or
transmitted. The average occupation number of the transmitted state
and reflected state are $ \langle n_T \rangle = T$ and $ \langle
n_R\rangle = R$. The fluctuations $\Delta n_{T} = n_{T}- \langle n_{T}
\rangle $ in the transmitted state and $\Delta n_{R} = n_{R}- \langle
n_{R} \rangle $ in the reflected state have mean squared fluctuations
and correlations \cite{mb92} given by
\begin{equation}
\langle (\Delta n_{T})^{2} \rangle  = 
\langle (\Delta n_{R})^{2} \rangle  = 
- \langle \Delta n_{T} \Delta n_{R} \rangle 
= RT.
\end{equation}
The fluctuations are maximal for $T = 1/2$ and vanish in the limit of
perfect transmission or complete reflection. The quantum partition
noise \cite{mb90,khlus,lesovik,tmrl} was observed in quantum point
contacts \cite{noiseqpc}, metallic diffusive wires and chaotic
cavities and other systems \cite{ybmb}.
\begin{figure}[top]
\centerline{\psfig{figure=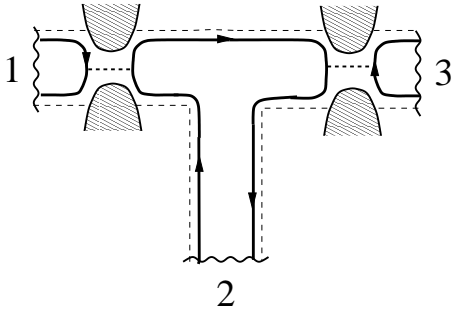,width=6.0cm}}
\caption{Partition noise geometry with two quantum point contacts.
The correlation between contacts $2$ and $3$ is measured. 
The left most quantum point contact permits to control the population 
of the edge state.}
\label{fig2}
\end{figure}
The above consideration applies only to single particles. Statistical
effects are a consequence of identical indistinguishable
particles. Thus an approach to noise is necessary which takes the
basic symmetries of many particle wave functions into account. In
second quantization a general formulation of the noise power for
non-interacting particles in terms of the scattering matrix
$s_{\alpha\beta}$ was given in Refs. \cite{mb90,mb92,mb91}.  The
scattering matrix $s_{\alpha\beta}$ relates current amplitudes
incident in contact $\beta$ to out-going current amplitudes in contact
$\alpha$.  In the zero-temperature limit the current-correlations of
interest are determined by a matrix
\begin{equation}
B_{\gamma\delta} = \sum_{\alpha}s_{\gamma\alpha}s^{\dagger}_{\delta\alpha}
(f_{\alpha} -f_{0}), 
\end{equation}
where $f_{0}$ is an arbitrary energy dependent function. 
In terms of this matrix we find
the cross-correlations ${\gamma \ne \delta}$,
\begin{equation}
S_{\gamma\delta} = - 2 (e^{2}/h) \int dE Tr[B^{\dagger}_{\gamma\delta}B_{\gamma\delta}].
\label{corr}
\end{equation}

This proofs that cross-correlations for Fermions are negative for
conductors in zero-impedance external circuits \cite{mb91,mb92}.
Would we repeat the calculation for photons emitted by black-body
radiators \cite{mb92} we would find that bunching of photons permits
positive correlations.  In what follows we use this expression to
evaluate the cross-correlations.
\begin{figure}[top]
\centerline{\psfig{figure=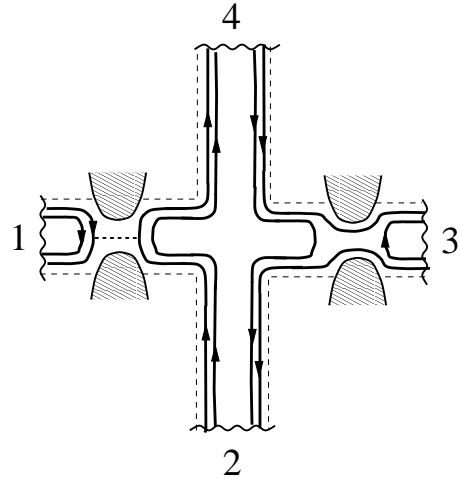,width=6.0cm}}
\caption{Partition noise geometry with two quantum point contacts.
The inner edge state is completely reflected at both QPC's. A
dephasing voltage probe (contact $4$) generates noise even if also the
outer edge state is perfectly transmitted at both QPC's.  An inelastic
voltage probe (contact $4$) converts the shot noise generated at the
left QPC into a positive current-correlation $S_{23}$.}
\label{fig3}
\end{figure}
\section{Partition noise experiments}
Consider the geometry of a conductor with two QPC's in series as shown
in Fig. \ref{fig2}, studied in the experiment of Oberholzer et
al. \cite{ober}. The experiment measured the
correlation $S_{23}$. Contact 1 is at a potential $eV$ and contacts
2 and 3 are grounded.  The left and right QPC's ($i =1$ and $i =2$)
are described by scattering matrices
\begin{equation}
\left( \begin{array}{cc} \sqrt{R_{i}} & \sqrt{T_{i}} \\ 
-\sqrt{T_{i}} &\sqrt{R_{i}} 
\end{array} \right) .
\label{smat1}
\end{equation}
The resulting zero-temperature correlation function is 
\begin{equation}
S_{23} = - 2 (e^{2}/h) |eV| T_{1}^{2} T_{2} R_{2} .
\label{oberholzer} 
\end{equation}
The transmission probability $T_1$ takes the role of the Fermi
function: for $T_1 = 1$ the edge state is fully filled (degenerate),
as $T_1$ becomes very small the edge state is only sparsely populated.
In this latter limit classical Maxwell-Boltzmann statistics applies
and the correlation function vanishes.
\section{Effect of dephasing and inelastic scattering}
Texier and one of us \cite{ctmb} investigated the effect of elastic
(interedge state) scattering, of dephasing and of inelastic
transitions on the correlation function given by
Eq. (\ref{oberholzer}).  Inelastic scattering and dephasing can be
investigated with the help of an additional contact, shown in
Fig. \ref{fig3}.  A real voltage probe acts as an inelastic scatterer.
The electrochemical potential $\mu_4$ fluctuates as a function of time
to maintain the total current into the probe at zero.  In contrast a
dephasing contact \cite{djcb,vlmb} preserves the energies of the
carriers at the contact: it is required that the current into the
dephasing contact vanishes at each energy.
\begin{figure}[top]
\centerline{\psfig{figure=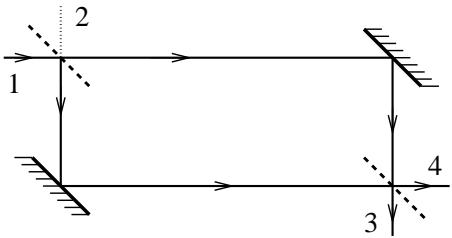,width=6.0cm}}
\caption{Optical Mach-Zehnder interferometer}
\label{fig4}
\end{figure}
For the single edge state used in the experiment by Oberholzer et al.
it turns out that the correlation Eq. (\ref{oberholzer}) is neither
sensitive to elastic nor inelastic scattering. But the question
becomes interesting if as shown in Fig. \ref{fig3} there are two edge
states.

The most interesting feature of the {\it dephasing} contact arises if
both QPC's transmit the outermost edge perfectly $T_1 = T_2 = 1$ and
reflect the innermost edge perfectly.  If the dephasing probe is
closed, the corresponding completely quantum coherent sample is
noiseless.  If we now switch on the dephasing voltage probe and allow
as shown in Fig. \ref{fig3} both edge channels to enter, the dephasing
voltage probe generates noise!! The distribution function $\bar f_4
(E)$ in the dephasing voltage probe, under the biasing condition
considered here, is still a {\it non-equilibrium} distribution
function and given by $\bar f_4 (E) = \frac{1}{2} f_1(E) + \frac{1}{2}
f_2(E)$.  The distribution function at the dephasing contact is
similar to a distribution at an elevated temperature with $kT =
|eV|/4$ with $eV$ the voltage applied between contact $1$ and contacts
$2$ and $3$.  We have $\int dE \bar f_4 (E) (1 - \bar f_4 (E)) =
e|V|/4$.  Evaluation of the correlation function gives \cite{ctmb},
\begin{equation}
\label{Sqe}
S^{\rm qe}_{23}= -({e^2}/{h})|eV| ({1}/{4}) .
\end{equation} 

The electron current incident into the voltage probe from contact $1$
is noiseless. Similarly, the hole current that is in the same energy
range incident from contact $2$ is noiseless. However, the voltage
probe has two available out-going channels. The noise generated by the
voltage probe is thus a consequence of the partitioning of incoming
electrons and holes into the two out-going channels. In contrast, at
zero-temperature, the partition noise in a coherent conductor is a
purely quantum mechanical effect. Here the partitioning invokes {\it
no quantum coherence} and is a {\it classical} effect \cite{rev}.
\begin{figure}[top]
\centerline{\psfig{figure=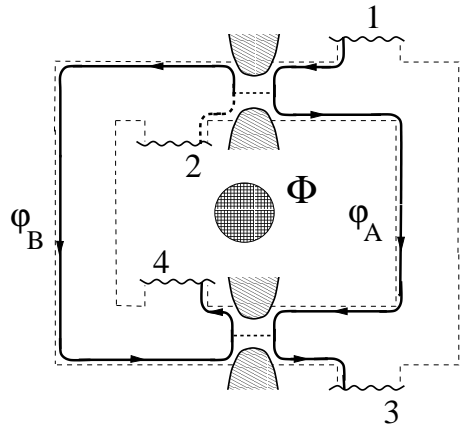,width=6.0cm}}
\caption{Edge state equivalent of the optical Mach-Zehnder interferometer.
An Aharonov-Bohm fux penetrates the hole of the structure.}
\label{fig5}
\end{figure}
The interesting effect in the presence of a {\it real} voltage probe
arises from the fact that the voltage in contact $4$ must fluctuate to
maintain the current at zero. Here we now permit the left contact to
have transmission probability $T_1 \le 1 $.  The resulting fluctuating
voltage leads to the possibility of correlated injection of carriers
into the two edge channels leaving the voltage probe.  As a
consequence the correlation function can now become positive even in a
purely normal conductor. For the geometry of Fig. \ref{fig3} we find
\cite{ctmb},
\begin{equation} 
\label{posi} 
S^{\rm in}_{23} =  + ({e^2}/{h})|eV| ({1}/{2}) T_{1} R_{1} . 
\end{equation}
This simple example shows that interactions can play an important role
and lead to surprising results. Other related problems which show such
a sign reversal include tunneling into a Luttinger liquid \cite{crep},
dynamical spin-blockade in a ferro-magnetic-lead normal-dot system \cite{cott}, 
or hybrid systems with superconductors, or frequency dependent
transport \cite{rev}. 
\begin{figure}[top]
\centerline{\psfig{figure=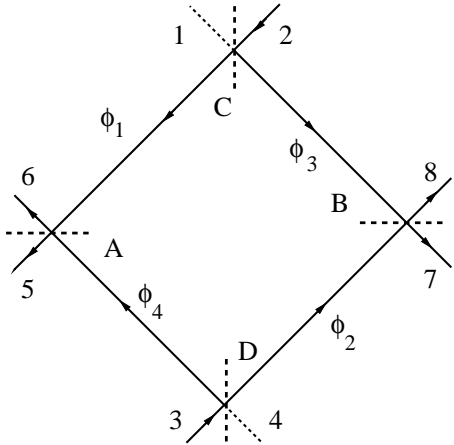,width=6.0cm}}
\caption{Two-source, four-detector optical Hanbury Brown Twiss geometry.
$\Phi$ represents an Aharonov-Bohm flux. 
After Ref. \protect\cite{sam2}.}
\label{fig6}
\end{figure}
\section{Mach-Zehnder geometry}
The optical Mach-Zehnder geometry is shown in Fig. \ref{fig4}.  The
central property of a Mach-Zehnder geometry is that carriers (photons)
exhibit only forward scattering.  This is in contrast with the typical
ring like structures used in mesoscopic physics where connection of a
lead to a ring generates invariably backscattering and closed orbits.
In principle, however, even in zero magnetic field a Mach-Zehnder
interferometer can be realized with the help of X-junctions (see
Ref. \cite{seelig}).  In high magnetic fields, with the help of edge
states, a Mach-Zehnder interferometer has recently been realized by Ji
et al. \cite{ji} and used to investigate the effect of decoherence on
the shot noise \cite{ji,marq}.  The structure is schematically shown
in Fig. \ref{fig5}. A Corbino like geometry is used with two QPC's.
An electron incident from contact 1 is at the first QPC either
transmitted or reflected. The transmitted partial wave proceeds along
the outer edge to the second QPC whereas the reflected partial wave
proceeds via the inner edge to the second QPC.

To be specific we choose the scattering matrix of the QPC to be given
by Eq. (\ref{smat1}).  To simplify the results we take $T = R = 1/2$.
The interference of the partial waves in the exiting channel
(second-order interference) leads to scattering matrix elements which
are functions of both phases $\phi_A$ and $\phi_B$. For instance
\begin{equation}
s_{31} = (1/2)[\exp(i(\phi_{B} - \psi_{2})) + \exp(i(\phi_{A}+ \psi_{1}))]
\end{equation}
Here we have in addition to the geometric phases $\phi_A$ and $\phi_B$
added the effect of an Aharonov-Bohm flux through the center of the
interferometer, $\psi_{1} + \psi_{2} = 2\pi \Phi/\Phi_{0}$ with
$\Phi_{0} = h/e$ the elementary flux quantum.

If contact $1$ is at an elevated potential $eV$ and all other contacts
are grounded the current at contact $3$ is determined by the
conductance $G_{31} = - (e^{2}/h) T_{31}$ which is

\begin{equation} 
G_{31} = - (e^{2}/2h) (1+ \cos(\phi_{B} - \phi_{A}- 2\pi \Phi/\Phi_{0}))
\end{equation}
The phase-dependence of the transmission and conductance is a
consequence of the superposition of partial waves in an out-going
channel.  Our goal is now to show that there are geometries in which
interference effects can arise in correlations even so all conductance
matrix elements are phase-insensitive.
\begin{figure}[top]
\centerline{\psfig{figure=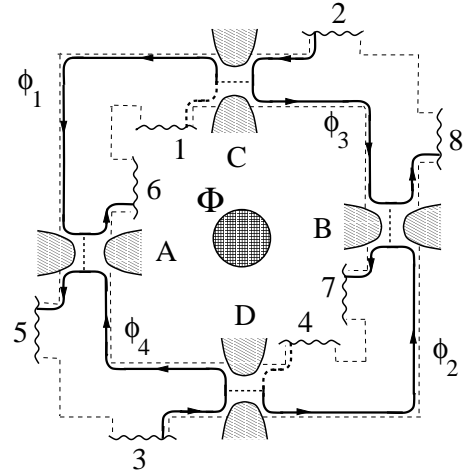,width=6.0cm}}
\caption{Two-source, four detector electrical Hanbury Brown Twiss geometry.
After Ref. \protect\cite{sam2}. 
}
\label{fig7}
\end{figure}

\section{Exchange interference in HBT experiments}
An optical two-source configuration \cite{yust} is shown in
Fig. \ref{fig6}.  It is equivalent to the stellar interferometer
experiment of HBT. Two sources $2$ and $3$ illuminate detectors at
contacts $5$ to $8$. Note that in this geometry there is no
interference due to splitting of an incident wave and superposition of
the resulting partial waves in an outgoing channel as in the
Mach-Zehnder interferometer.  Sources $2$ and $3$ are incoherent and
their intensities at a detector add classically.  Nevertheless the
intensity-intensity correlations at contact pairs $58,57,68$ or $67$
are functions of the phases $\phi_1$ to $\phi_4$ as we will now show.

The electrical edge state equivalent \cite{sam2} of the optical
geometry of Fig. \ref{fig6} is shown in Fig. \ref{fig7}. The elements
of the scattering matrix now contain phases only in a trivial
multiplicative way.  For instance the scattering matrix element for
transmission from source contact $2$ to the detector contact $5$ is
\begin{equation}
s_{52} =   T_{A}^{1/2} \exp(i\phi_1 ) T_{C}^{1/2}
\end{equation}
Here $T_{C}$ and $T_{A}$ are the transmission amplitudes of the QPC's
denoted $C$ and $A$ in Fig. \ref{fig7}. Similar expressions hold for
all other elements of the s-matrix.  In contrast to the Mach-Zehnder
interferometer here phase factors like $\exp(i\phi_1 )$ simply
multiply the scattering matrix elements. An Aharonov-Bohm (AB) flux
through the hole of this Corbino geometry similarly introduces only
multiplicative phase factors. Consequently the conductance matrix is a
function of the transmission and reflection probabilities of the
individual QPC's only.

To be definite let us take the transmission and reflection
probabilities at the QPC denoted $C$ to be $T_C=T$ and $R_C=R$ and at
$D$ to be $T_D=R$ and $R_D=T$.  If the source contacts $2$ and $3$ are
at a potential $eV$ and all other contacts are grounded the currents
at the detector contacts are \cite{sam2}
\begin{equation}
I_5 = I_6= ({2e^2}/{h}) TV, \hspace{0.5 cm} I_7 = I_8 = ({2e^2}/{h}) RV
\end{equation}
independent of the transmission amplitudes of the QPC's denoted by $A$
and $B$ in Fig. \ref{fig7}. Most importantly, the corresponding conductances
reveal no phase dependence and are independent of the AB-flux $\Phi$.

In contrast to the conductance matrix the correlation functions now
depend on the phases. Consider the simple case where transmission and
reflection amplitudes of all QPC's are equal to $1/2$. The correlation
function $S_{58}$ is \cite{sam2}
\begin{eqnarray}
S_{58} &=& - (e^{2}/4h) |eV| \nonumber \\
&\times&(1 + \cos (\phi_1 + \phi_2 - \phi_3 - \phi_4)). 
\label{hbtphases}
\end{eqnarray}
The correlation function thus depends in an essential way on the
phases accumulated by different particles from the source contact to
the detector contact. An AB-flux through the hole of the Corbino disk
contributes a positive phase to $\phi_1$ and $\phi_2$ and a negative
phase to $\phi_3$ and $\phi_4$ to give a total additional phase
contribution of $2\pi \Phi/\Phi_{0} $. Thus we have a geometry for
which conductances exhibit no AB-effect but correlation functions are
sensitive to the variation of an AB-flux!

The two particle Aharonov-Bohm effect demonstrates nicely that quantum
interference is not simply related to properties of single particle
states.  The existence of such an additional phase-sensitivity of
forth-order interference was recognized in early work on noise
correlations: An early paper \cite{mb91} on noise in multi-terminal
mesoscopic conductors was entitled "The quantum phase of flux
correlations in wave guides". However, neither in this work nor in
subsequent efforts \cite{mbprl92,gram,dles} was a geometry found in
which the effect can be seen in such a clear cut way as in the
geometry of Fig. \ref{fig7}. The first term in Eq. (\ref{hbtphases})
is the sum of the correlations that are obtained if only source $2$ is
active and if only source $3$ is active. These single source
correlations are phase-insensitive.  Indeed in the presence of
complete dephasing (described e.g. with the help of an energy
conserving voltage probe, say at the outer edge between QPC's $C$ and
$A$) it is precisely the first term in Eq. \ref{hbtphases} which
survives \cite{note}.

In optics, set-ups with independent sources similar to Fig. \ref{fig6}
have been proposed theoretically \cite{yust} and investigated
experimentally \cite{fran} in the context of entanglement and Bell
Inequalities(BI). However, non-thermal sources and coincidence
measurements have been employed. Neither of these are available in
electrical conductors. Nevertheless, the joint photon detection
probabilities \cite{bell} used to test BI, have the same phase
dependence as the correlation function Eq. (\ref{hbtphases}). This
naturally raises the question if the above correlations are not also
in fact a consequence of entanglement of the carriers emitted by the
two reservoirs? Moreover, if this is the case, can this entanglement
be used to violate a BI expressed in terms of the zero-frequency
correlators? Below, we show that the answer to both these questions is
yes.
\section{Entanglement and HBT experiments}
Instead of modulating the phases in Eq. (\ref{hbtphases}) via the path
lengths, we investigate the correlation functions in the two-source
HBT set-up of Fig. \ref{fig7} by varying the transmission through the
two QPC's $A$ and $B$ which precedes the detector contacts. This is
similar to schemes in optics where one varies the transmission to the
detectors with the help of polarizers. The advantage of this latter
approach is that the Bell Inequalities in the presence of dephasing
\cite{sam1} (in general not a problem in optics) can be violated over
a wider range of parameters \cite{been2}.

Detection of entanglement via violation of a Bell Inequality (BI) in
electrical conductors is discussed in
Refs. \cite{mait,cht,sam1,been,fazio,gisin}. Following the original
suggestion of Bohm and Aharonov\cite{bohm}, it is most tempting to
treat spin entanglement. However, it is charge current fluctuations
that are measured and the conversion of spin to charge information
adds to experimental complexity. Entanglement can, however, take other
forms \cite{sam1,been,gisin} and below we follow
Refs. \cite{sam1,been} and discuss entanglement of orbital degrees of
freedom.

To simplify the following discussion we now assume an AB-flux $\Phi$
such that the phases in Eq. (\ref{hbtphases}) add up to a multiple of
$2\pi$.  The transmission and reflection probabilities through the
detector QPC's are taken to be $T_A = 1- R_A = \sin^{2}(\theta_A)$ for $A$
and with $\theta_A$ replaced by $\theta_B$ for $B$).  Ref. \cite{sam2}
finds the noise powers
\begin{eqnarray}
S_{58} = S_{67} = - ({2e^2}/{h})|eV|RT\cos^2(\theta_A-\theta_B), \\
S_{57} = S_{68} = - ({2e^2}/{h}) |eV|RT\sin^2(\theta_A-\theta_B).  
\label{noise1}
\end{eqnarray}
The non-local dependence on the angles $\theta_A$ and $\theta_B$ is
just the one found for the two-particle joint detection probability in
the context of two-particle entanglement, as originally discussed by
 Bell \cite{bell}.

To provide a clear picture of entanglement in the HBT geometry we now
explicitly construct the many-body state generated by the two
incoherent source reservoirs $2$ and $3$. In a second quantization
notation (suppressing the spin index) the transport state
generated by the two sources is
\begin{equation}
|\Psi \rangle=\prod_{0<E<eV}c_2^{\dagger}(E)c_3^{\dagger}(E)|0\rangle
\label{state}
\end{equation}
where $|0\rangle$ is the ground state, a filled Fermi sea in all
reservoirs at energies $E<0$. The operator $c^{\dagger}_{\alpha}(E)$
creates an injected electron from reservoir $\alpha$ at energy $E$. To
analyze this state consider first a pair of single electrons at energy
$E$ described by $c_2^{\dagger}c_3^{\dagger}|0\rangle$.  Let us denote
the creation operators of particles reflected at $C$ by
$c_{2B}^{\dagger}$ and of particles transmitted at $C$ by
$c_{2A}^{\dagger}$.  Similarly, let us denote the creation operators
of particles reflected at $D$ by $c_{3A}^{\dagger}$ and of particles
transmitted by $c_{3B}^{\dagger}$.  The second index $A,B$ thus
denotes towards which beam-splitter the electron is propagating. The
state (keeping in mind that $C$ transmits with probability $T$ and $D$
transmits with probability $R$) beyond the QPC's C and D is then
\cite{sam2}
\begin{equation}
|\Psi \rangle = |\tilde \Psi\rangle + |\Psi_T\rangle + |\Psi_R\rangle
\end{equation}
with 
\begin{eqnarray}
|\tilde \Psi \,\,\,\rangle &=& \sqrt{RT}
\left(c_{2A}^{\dagger}c_{3B}^{\dagger} + 
c_{2B}^{\dagger}c_{3A}^{\dagger}\right)|0\rangle
\nonumber \\ 
|\Psi_T\rangle &=& Tc_{2A}^{\dagger}c_{3A}^{\dagger}|0\rangle ,
|\Psi_R\rangle = Rc_{2B}^{\dagger}c_{3B}^{\dagger}|0\rangle
\label{twostates}
\end{eqnarray}
The total state $|\Psi\rangle$ consists of a contribution, $|\tilde
\Psi\rangle$, in which the two particles fly off one to $A$ and one to
$B$, and of two contributions, $|\Psi_R\rangle$ and $|\Psi_T\rangle$,
in which the two particles fly both of towards the same detector
contact.

A particularly simple limit is the case of strong asymmetry $R \ll 1$,
where almost all electrons are passing through both source beam
splitters towards detector $A$.  In this limit, the state
$|\Psi_R\rangle$ can be neglected.  Moreover, by redefining the vacuum
to be the completely filled stream of electrons, i.e. introducing
$|\bar 0 \rangle=c_{2A}^{\dagger}c_{3A}^{\dagger}|0\rangle$, we can
write the state $|\Psi \rangle$ to leading order in $\sqrt{R}$ as
\begin{eqnarray}
|\Psi \rangle=\left(1+\sqrt{R}\left[c_{3B}^{\dagger}c_{3A}-
c_{2B}^{\dagger}c_{2A}\right]\right)|\bar 0\rangle 
\label{newgs}
\end{eqnarray}
The operators $c_{3A}(E)$ and $c_{2A}(E)$ now describe hole
excitations, i.e. the removal of electrons from the filled stream. To
leading order in $\sqrt{R}$ the total state in Eq. (\ref{state}),
including again the full energy dependence, can thus be written as
\begin{eqnarray}
&|\Psi \rangle & = |\bar 0 \rangle + 
\nonumber \\
&\sqrt{R}& \int_0^{eV}dE
 \left[c_{3B}^{\dagger}(E)c_{3A}(E)-c_{2B}^{\dagger}(E)c_{2A}(E)\right]|\bar
 0\rangle 
\label{stateexp}
\end{eqnarray}
Due to the redefinition of the vacuum \cite{sam1}, we can interpret
the resulting state as describing a superposition of electron-hole
pair excitations out of the ground state (created at C and D), i.e. an
orbitally entangled pair of electron-hole excitations. This is
equivalent to the recent findings by Beenakker et al, \cite{been}, who
discussed the generation of entangled electron-hole pairs at a single
QPC. The state is similar to the two-electron state 
considered by
Samuelsson, Sukhorukov and B\"uttiker, \cite{sam1}, emitted from a
superconductor contacted at two different points in space. We note
that that the new vacuum $|\bar 0\rangle$ does not contribute to the
current correlators since it describes a filled, noiseless stream of
electrons.

Can we formulate a BI in terms of the zero-frequency cross-correlators
Eq. (\ref{noise1})? In the strongly asymmetric case, $R \ll 1$, this
is clearly the case. The state in Eq. (\ref{stateexp}), describes
Poissonian emission of orbitally entangled electron-hole pair wave
packets. Following Ref. \cite{sam1}, the average time between each
emission is much longer than the coherence time of each pair, and the
zero frequency cross correlations are just identical to a coincidence
measurement, running over a long time. The electron-hole joint
detection probability is proportional to the zero frequency cross
correlations. One can then apply the arguments in the original paper
of Bell \cite{bell} and directly formulate a BI in terms of the zero
frequency cross correlators \cite{sam1} for four different angle
configurations $-2\leq S_{B} \leq 2$,
in the form \cite{clau,asp}
\begin{equation}
S_{B} = E(\theta_A,\theta_B) - E(\theta_A,\theta_{B}') + 
E(\theta_{A}',\theta_B) + E(\theta_{A}',\theta_{B}')
\label{BI}
\end{equation}
where the Bell correlation functions
\begin{equation}
E(\theta_A,\theta_B)= {\left(S_{58}+S_{67}-S_{57}-S_{68}\right)}/S_{0}
\label{corrfunc2}
\end{equation}
with $S_{0} = - {(4e^2RT/h)}|eV|$ are given by 
\begin{equation}
E(\theta_A,\theta_B) = \cos(2[\theta_A-\theta_B]) .
\label{corrfunc3} 
\end{equation}
By adjusting the four angles the maximal Bell parameter is $S = 2
\sqrt{2}$ and the BI is thus violated.

We note that the phase dependence of the current correlators in
Eq. (\ref{noise1}) is not a result from taking the limit $R\ll
1$. However, for $R$ not small, many electron-hole pairs are
superimposed on each other during the measurement, i.e. the average
emission time of the electron-hole pairs is of the order of the
pair-coherence time. As pointed out above, the derivation of the BI in
Eq. (\ref{BI}) rest on the assumption that the pairs are well
separated, and the BI can thus not be applied for arbitrary
$R$. Whether the phase dependence of the correlators in
Eq. (\ref{noise1}) results from some kind of multiparticle entanglement
and whether this entanglement can be detected via a violation of a BI,
are interesting questions which however go beyond the scope of the
present paper.

We have discussed only the case of integer quantum Hall states. The
fractional quantum Hall effect offers a wider, very interesting, area
for the examination of correlations
\cite{safi,vish} since in this case fractional statistical effects are
realized.

The HBT effect and the BI both concern two particle effects.
In this work we have established a relation between the two.  
The simple adiabatic edge-state geometry described 
above using only normal reservoirs as particle sources,
the focus on orbital non-locality and the use of zero-frequency correlators 
brings experimental detection of entanglement in electrical conductors
within reach. 

\begin{ack}
This work was supported by the Swiss National Science Foundation
and the program for Materials with Novel Electronic Properties.
We have benefited from interactions with C. W. J. Beenakker and 
N. Gisin. \\
\end{ack}

\noindent{\bf Note added in proof}\\

Since submission of this work, we have included a treatment in
Ref. \cite{sam2} of entanglement and violation of a Bell Inequality
due to post-selection for a nearly symmetric (large transmission
interferometer) going beyond the tunneling limit. We acknowledge
discussion and correspondence (unpublished) with C. W. J. Beenakker on
this topic.


\begin{thebibliography}{9}
\bibitem{note}  A related question is the robustness of exchange interference 
                effects under ensemble averaging. It is shown Refs. \cite{vlmb,ybmb,sukh}
                that even if the contributions to the correlation from single 
                sources are subtracted ensemble averaging gives a finite 
                result.
\end{thebibliography}
\end{document}